\def\mag{\hbox{$\;.\!\!\!^m$}}
\def\to{\hbox{$\,$--$\,$}}
\def\muspc{\hskip 0.15 em}

\def\zdot{\hbox{${\rm z}\hskip -0.36em^{^{\fiverm\bullet}}\hskip -0.08em$}}
\hyphenpenalty=50

\input aa.cmmb
\pageno=1
\MAINTITLE{On the extinction towards Baade's Window}
\SUBTITLE{}
\AUTHOR{Y.K. Ng@{1} and G. Bertelli@{2,3}}
\INSTITUTE{
 @1
 \hbox to 10.5cm{
 IAP, CNRS, 98 bis Boulevard Arago, 75014 Paris, France 
 \hfill}
 (\hbox to 1.2cm{E-mail:\hfill}{\tt ng\char64 iap.fr})
 @2
 \hbox to 10.5cm{
 Department of Astronomy, Vicolo dell'Osservatorio 5, 35122 Padua, Italy
 \hfill}
 (\hbox to 1.2cm{\null\hfill}{\tt bertelli\char64 astrpd.pd.astro.it})
 @3
 \hbox to 10.5cm{
 National Council of Research, CNR -- GNA, Rome, Italy
 \hfill}
}

\DATE{Received 16 April 1996 / Accepted 5 May 1996}

\ABSTRACT{An analysis is made on the 
morphology of the sequences,
formed by stars 
distributed along the line of sight 
in (V,V--I) CMDs (Colour Magnitude Diagram)
from Baade's Window (\hbox{$l\!=\!\!1\fdg0,\ b\!=\!\!-3\fdg9$}).
The extinction is due to an absorbing layer with an
effective thickness of about 100 pc. A linearly, with 
distance, increasing extinction up to a maximum value
inside the absorbing layer
is in first order a good approximation for the actual extinction
present along the line of sight.\par
We show that the morphology of the disc sequence 
gives a good indication of the nearby extinction along the
line of sight.
The morphology of the bulge/bar red horizontal branch
is likely due to a
combination of differential reddening and a significant 
metallicity spread among the horizontal branch stars.
This is irrespective of the type of extinction law adopted.
There is in Baade's Window globally no big difference between a Poissonian 
or a patchy type of extinction,
except for substructures in the morphology of the sequences
for the latter.
}
\KEYWORDS{
methods: data analysis
 -- Stars: HR-Diagram
 -- Galaxy: stellar content, structure }
\THESAURUS{04(3.13.2, 08.08.1, 10.19.2, 10.19.3)
}
\maketitle

\SUBTITLE{On the extinction towards Baade's Window}
\MAINTITLE{Y.K. Ng \& G. Bertelli:}

\titlea{Introduction}
Synthetic Hertzsprung-Russell Diagrams (HRDs) can be
generated through the population synthesis technique,
with the aid of libraries of stellar evolution tracks
(Bertelli et al. 1994a) of different chemical compositions
These HR-diagrams are part of a powerful tool,
the HRD-GST (HRD Galactic Software
Telescope; see Ng 1994 \& 1996 and Ng et al. 1995 for details),
in stars counts studies of the Galactic Structure,
in particular the structure towards the Galactic
Centre. Using the HRD-GST, the contributions of various
stellar populations are decomposed
statistically. Through a detailed and a concise
analysis of the star counts along the line of sight we aim to study

\item{$\bullet$} the galactic structure;
\item{$\bullet$} the interstellar extinction; and
\item{$\bullet$} the ages and metallicities of the different 
stellar populations.

\noindent
The results from our studies are 
reported in the papers by Ng et al. (1995 \& 1996a), Bertelli
et al. (1995), these papers will be hereafter referred to as 
Paper I--III, and Bertelli et al. (1996) \& Ng et al. (1996b).
\par
The ages, metallicities and the spatial distributions of the stars from
various populations contain a wealth of information
about the formation and evolution of our Galaxy.
The stars, observed along any line of sight, are the result of
a complex mixture from various populations. Especially towards the
Galactic Centre there appears to be no clear consensus about
the ages \& metallicities of the stellar populations
(disc, bulge, bar ...)
and the parameterization of the galactic structure.
The large variation of the extinction over a relative
small area is one of the major causes.
The near-IR passbands are less sensitive to extinction.
%but at the same time they are also less sensitive for
%metallicity variations.
Each field needs to be studied separately,
because the extinction over the various fields is not
easily parameterized.
The best strategy is probably, to determine the extinction 
(if this is not too high and/or patchy)
\& the age-metallicity of the various stellar populations 
from the optical passbands and verify the results with
near-IR photometry.
\par
In (V,V--I) CMDs extinction and age-metallicity effects are difficult to 
separate from each other. 
The tilted clump of HB (Horizontal Branch) stars 
could be due to large differential extinction, 
a large metallicity spread of the HB stars, or a 
combination of both
(Catalan \& de Freitas Pacheco 1996; Ortolani et al. 1990 \& 1995a; 
Ng et al. 1996a\&d).
This might imply, as suggested by Renzini (1995),
that there are no super metal-rich stars towards the Galactic Centre.
Patchy extinction might explain the various structures present in the
(V,V--I) CMDs. 
In addition the metal-rich globular clusters might be 
related to the bulge (Ortolani et al. 1995a)
or the bar (Ng et al. 1996d)
in both age \& metallicity.
Establishment to which population these clusters are related 
with, would provide valuable information about part of the 
formation history of our present day Galaxy.
\par
\begfig 9.9cm
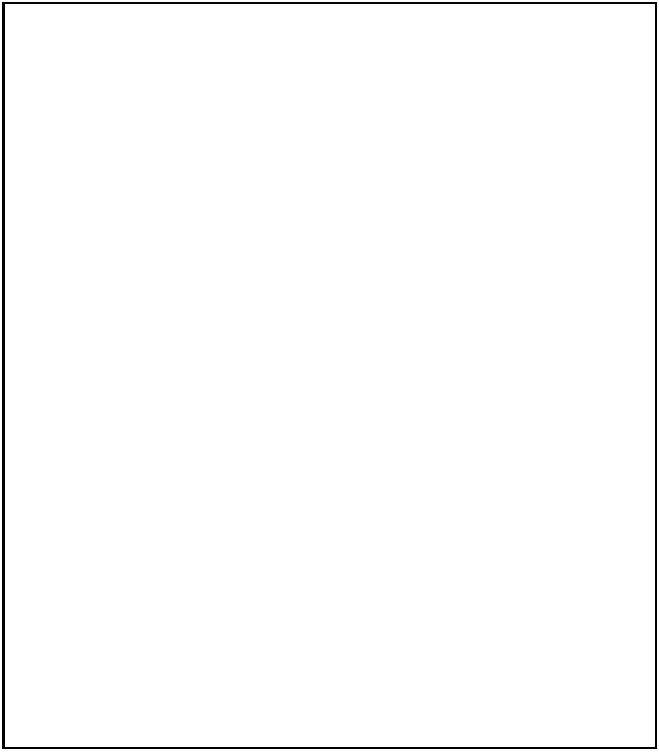
\figure{1.1}{(Synthetic (V,V--I) CMD for Baade's Window.
For sake of simplicity not all populations have been inserted.
Only the disc populations with a scale height less than 250 pc
and the `bar' population (Ng et al. 1996a\&c) are displayed.
Dots, dashes, open circles and filled squares refer
respectively to main sequence stars, core H-exhausted or red giant branch
stars, horizontal branch stars and asymptotic giant branch stars}
\endfig
The objective of this paper is
to show, how the HRD-GST deals with extinction and 
to analyze the influence of the 
various manifestations of the extinction.
Emphasis is put on the extinction 
in the optical passbands and on the morphological appearance
of the structures in the CMDs. 
We show what the effects are
for a specific population. Baade's Window is used as an example.
In contrast to Stanek (1996), we show which
signatures can be expected in (V,V--I) CMDs
from different types of extinction. 
Combination of the different morphological 
structures in the CMDs allows us to determine the 
extinction along the line of sight, see 
Bertelli et al. (1995) and Ng et al. (1996a\&d).
In Sect. 2 we describe our method, show the results and discuss
them in Sect. 3. We end this paper with a summary of 
the results.
\par
\begfig 9.9cm
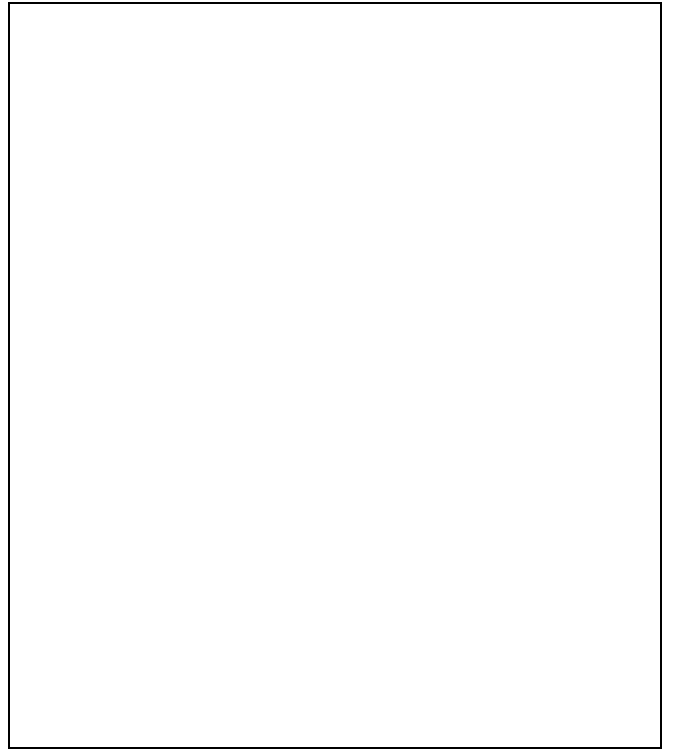
\figure{1.2}{Synthetic (V,V--I) CMD for Baade's Window
in absence of extinction.
For sake of simplicity not all populations have been inserted.
Only the disc populations with a scale height less than 250 pc
and the `bar' population (Ng et al. 1996a\&c) are displayed}
\endfig

\titlea{Extinction}

\titleb{Method}
As starting and reference point we will use the extinction obtained 
from our analysis (Ng et al. 1996a) of Baade's Window
CMD (Paczy\'nski et al. 1994 and references cited therein). 
We use the extinction law
E$_{V-I}$\muspc=\muspc A$_{V}$\muspc/2.4 to determine
the extinction in the I passband.
The analysis of 
Wo\'zniak \& Stanek (1996) justifies that this is a valid
approach in Baade's Window.
Figures~1.1 \& 1.2 show 
the simulated CMDs for Baade's Window field, respectively 
with and without extinction along the line of sight.
Figure 1.2 shows the change of the morphology in the CMD
without extinction, indicating that the disc sequence 
provides useful indications about the nearby extinction
along the line of sight.
In this case almost
all the structures
move to bluer colours and 
brighter magnitudes. The disc sequence becomes a vertical structure,
because we observe stars with the same colours at different distances. 
The dispersion is partly due to the photometric errors \& the 
crowding simulation
and partly due to age-metallicity differences among the disc stars. 
This figure also shows that,
without extinction, the detection of the 
bar's main sequence turn-off could have been possible.
\par
For sake of simplicity and clarity of the figures,
in which the sensitivity is demonstrated to
the foreground extinction,
we considered only the foreground stars
(i.e. populations with a scale height of 
250~pc or less,
say \hbox{d\muspc$<$\muspc4 kpc}), together with the allegedly called 
`bar' population. 
The contribution from other populations 
(old disc, thick disc, bulge, halo; see Ng et al. 1996b for 
a description of these population)
have not been considered.
The symbols in the figures designate approximately
the evolutionary phase of each synthetic star.
The extinction along the line of sight is varied in the following 
way:
\par
\begfig 5.5cm
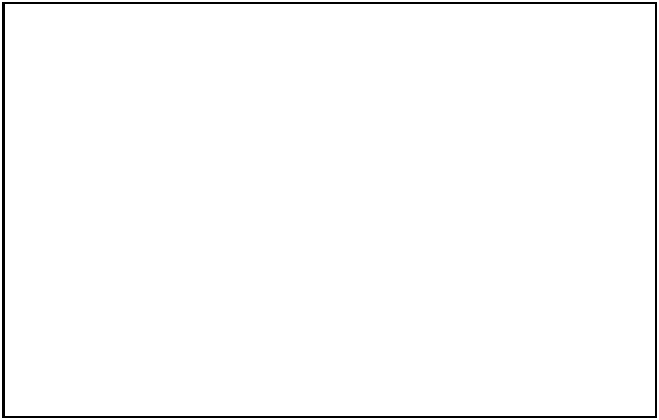
\figure{2ab}{The extinction along the line of sight in Baade's Window
for various linearly (long-dashed line) and step-like (dashed line)
increasing extinction curves,
at respectively 1, 2 and 4 kpc away from us.   
The thick solid line is the mean extinction for subfield \#3
(Ng et al. 1996a)}
\endfig
\begfig 5.5cm
%\begfig 5.5cm
%\psfig{file=fig2c.ps,height=5.5cm,width=8.6cm}
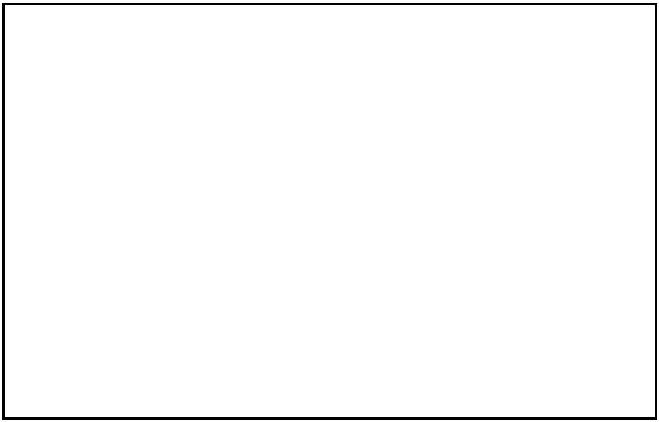
\figure{2b}{The extinction along the line of sight in Baade's Window
for a Poissonian distribution
(A$_{\rm V}$\muspc=\muspc1\mag60\muspc$\pm$\muspc0\mag20). 
The dots show the extinction
from the synthetic stars `detected' in the Monte-Carlo simulation.
The thick solid line is the mean extinction for subfield \#3
(Ng et al. 1996a)}
\endfig
%
%
%\begfigps 5.5cm
%\begfig 5.5cm
%\psfig{file=fig2b.ps,height=5.5cm,width=8.6cm}
%\figure{2b}{The extinction map along the line of sight in Baade's Window
%for various step-like extinction curves.
%The dotted line, dashed and solid lines
%designates the curves for which the change in extinction 
%occurs at respectively 1, 2 and 4 kpc away from us.   
%The thick solid line is the mean extinction 
%for subfield \#3 (Ng et al. 1996a)
%\vskip -3.0mm
%}
%\endfig
%
%
\begfig 5.5cm
%\begfig 5.5cm
%\psfig{file=fig2d.ps,height=5.5cm,width=8.6cm}
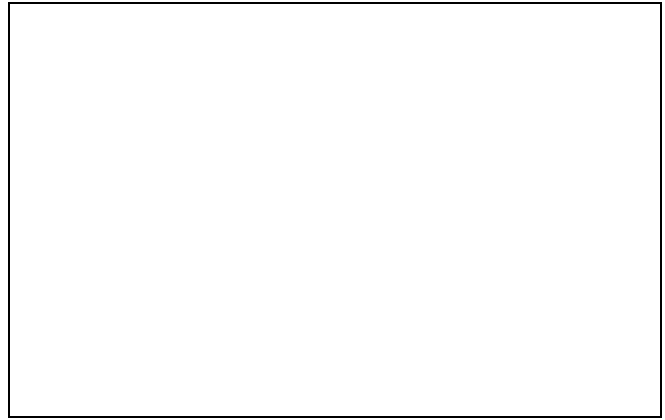
\figure{2c}{The extinction along the line of sight 
for a patchy ($\lambda$\muspc=\muspc0.6; 
\hbox{A$_{\rm V}$\muspc=\muspc0\mag0\to1\mag75}) 
distribution. Exaggerated, in order to enhance the
contrast of the lumpiness.
The dots show the extinction
from the synthetic stars `detected' in the Monte-Carlo simulation.
It is emphasized that this does not represent the 
actual situation in Baade's Window. 
The thick solid line shows the mean extinction for Baade's Window 
subfield \#3 (Ng et al. 1996a)}
\endfig
\item{$\circ$} use a linear relation for the extinction curve 
between us and 1, 2 or 4 kpc; the extinction is 
taken equal to maximum value, A$_{\rm V}$\muspc=\muspc1\mag75, at the latter
distances;
\item{$\circ$} use a step function for the extinction curve and increase
the extinction to A$_{\rm V}$\muspc=\muspc1\mag75 at 
1, 2 or 4 kpc away from us;
\item{$\circ$} use a random extinction irrespective of the
distance of the synthetic star along the line of sight; the distribution 
of the extinction is  
\itemitem{a)} Poissonian, or
\itemitem{b)} patchy.

\par
\noindent
In the Poissonian case we assume the presence of 
an absorbing layer. For the effective thickness of this layer 100 pc 
(Cox 1989) is adopted.
With an exponential density profile (Jones et al. 1981, Ruelas-Mayorga 1991)
the extinction will be noticeable up to three
times the effective thickness of this layer.
We therefore use 300 pc as the boundary of this absorption layer.
Outside this layer the extinction is equal to the maximum value.
Inside this layer the extinction is scaled 
linearly with its distance.
Arp (1965) showed that this is a valid
approach for Baade's Window.
This interpolation will suffice for the current purpose,
see for example Figs. 2b \& 2c.
\par
Several cases are considered for both the Poissonian and
the patchy extinction.
In the first case, the extinction is randomized between 
\hbox{A$_{\rm V}$\muspc=\muspc0\mag0\to1\mag75}
and in the second case 
\hbox{A$_{\rm V}$\muspc=\muspc1\mag60\muspc$\pm$\muspc0\mag20} is adopted.
The first range, see Fig~2c, is not representative for the
actual extinction in Baade's Window, because there are no regions where 
extinction is absent. It has been considered in order to exaggerate 
the difference between Poissonian and patchy extinction and 
furthermore, to enhance the contrast between the extinction clumps.
The latter value covers approximately the range
of mean extinction values determined for the Baade's Window 
subfields (Paczy\'nski 1994, Ng et al. 1996a).
For this part of the analysis 
the low and high extinction patches determined by 
Stanek (1996) are ignored.
In first approximation, 
\hbox{A$_{\rm V}$\muspc=\muspc1\mag30\muspc\to\muspc2\mag30}
is considered as a reasonable range 
for the total extinction range in Baade's Window.
\par
%***
%Stanek did not take the effects of 
%differences in metallicity
%of the clump stars into account for
%the determination of the extinction map.
%This would have been a reasonable approach if a high metallicity clump
%star does not trace a higher extinction in the analysis.
%But high metallicity clump stars are located in the CMD in the direction
%where a higher extinction is expected. His upper extinction
%limit might be slightly too high. Due to metallicity effects
%the upper limit of the extinction might be overestimated by 
%about 0\mag5.
%\par
A patchy extinction map is made with an algorithm,
which generates N$_1$ clustered points around 
N$_0$ Poissonian distributed parent points. The N$_1$ points are 
uniformly distributed within a radius $\lambda$.
This latter parameter defines the size of the extinction patch
and is expressed in units of the mean distance between the 
centre positions of the patch centers.
N$_0$ and N$_1$ define the resolution of the extinction map.
Many different realizations can be made for different
choices of the parameters. For our current purpose we
made runs with relatively small \& big patches by using
$\lambda\!=\!0.3$\ and $\lambda\!=\!0.6$.  
\par
Figures 2b \& 2c show respectively, the Poissonian extinction and
the patchy extinction. 
We emphasize that these figures show the total and not the
differential extinction along the line of sight. Each dot represents 
the extinction assigned to a synthetic star in the simulations.
Note that the extinction 
becomes clearer at larger distances, due to an increase 
in the detection 
of stars, which are located farther away 
in a larger volume. 
\par
\noindent
The large number of stars from the bar population
gives a high concentration of stars at 7\to9~kpc 
distance.
Note however, that this concentration is located at
a mean distance slightly larger than 8~kpc. While 8~kpc (Wesselink 1987)
has been adopted in the HRD-GST 
for the distance to the Galactic Centre (Ng et al. 1995).
The displacement of the mean concentration of stars beyond 8~kpc
is due to the fact that the 
differential volume increases with the distance.
\par
\begfig 9.9cm
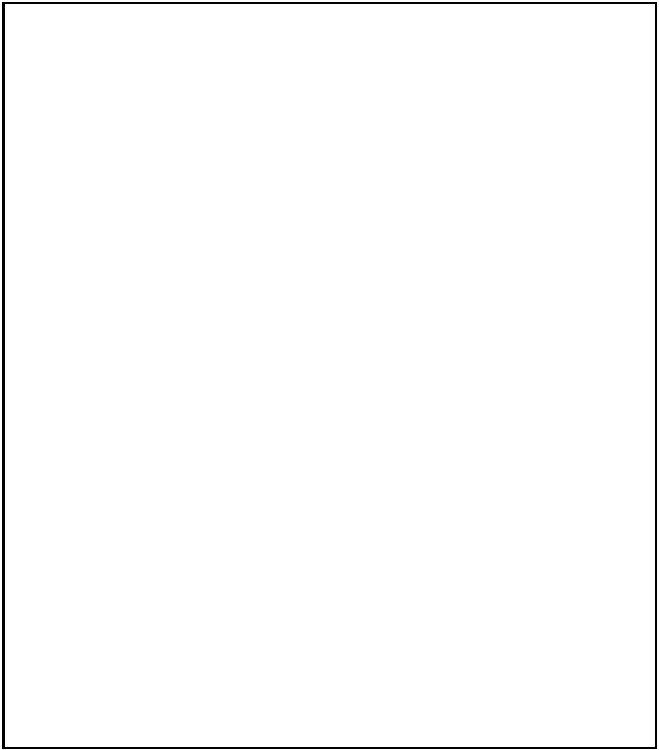
\figure{3.1}{Synthetic (V,V--I) CMD 
with a linearly increasing extinction up to A$_{\rm V}$\muspc=\muspc1\mag75
at 1 kpc distance
}
\endfig
Figures 2a\to2c show the different forms of
extinction adopted for this analysis. In all these figures we show,
as a reference,
the `actual' extinction (Ng et al. 1996a) as a thick line.
Figure 2a shows the linear and step-like extinction.
We consider three cases:
the extinction is increased linearly or stepwise to
the maximum value \hbox{A$_{\rm V}$\muspc=\muspc1\mag75} at a distance
1, 2 or 4 kpc away from us.
\par
\begfig 9.9cm
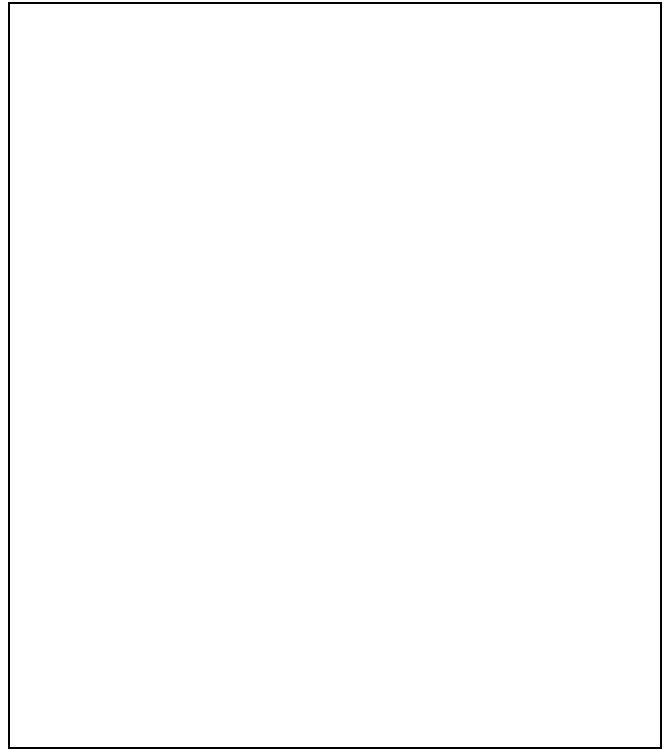
\figure{3.2}{Synthetic (V,V--I) CMD 
with a linearly increasing extinction up to A$_{\rm V}$\muspc=\muspc1\mag75
at 2 kpc distance
}
\endfig
\begfig 9.9cm
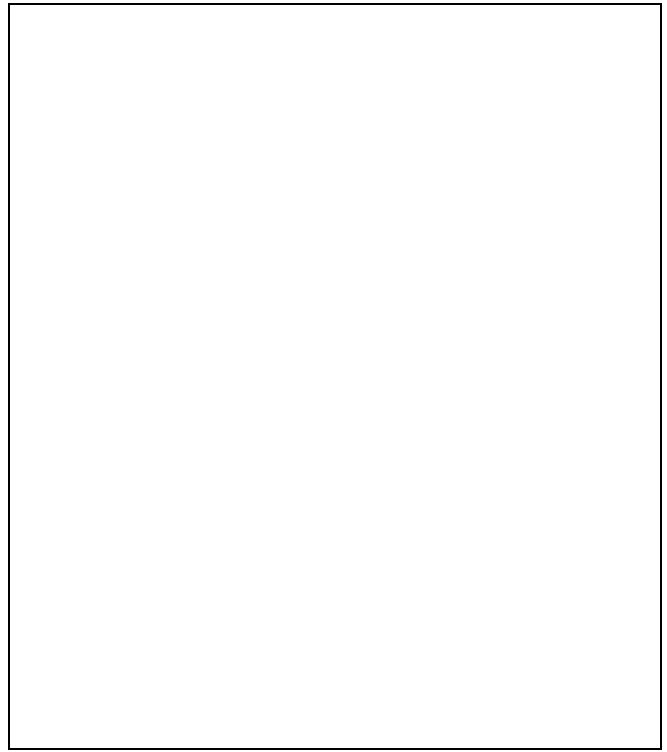
\figure{3.3}{Synthetic (V,V--I) CMD 
with a linearly increasing extinction up to A$_{\rm V}$\muspc=\muspc1\mag75
at 4 kpc distance
}
\endfig
%
%
%\vfill\eject
%\newcolumn
%\null
%
\begfig 9.9cm
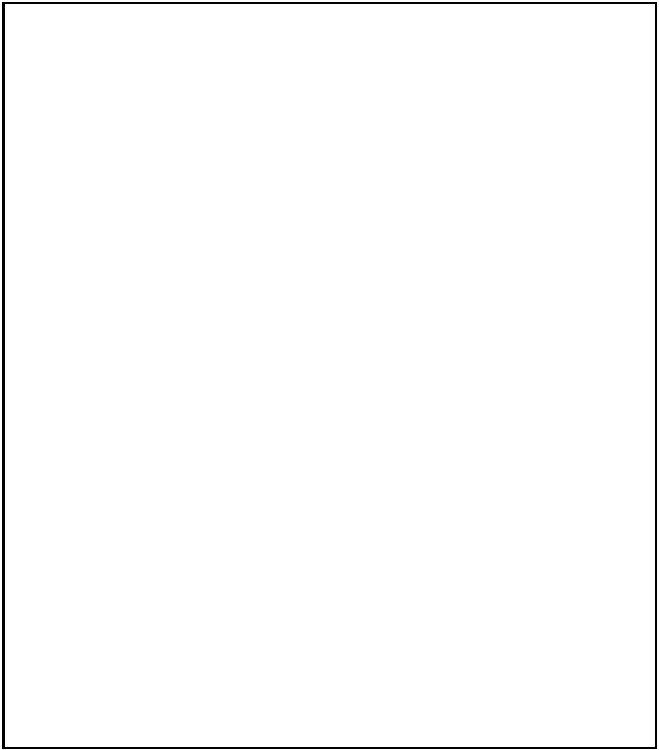
\figure{4.1}{Synthetic (V,V--I) CMD 
with a discretely increasing extinction up to A$_{\rm V}$\muspc=\muspc1\mag75
at 1 kpc distance
}
\endfig
\begfig 9.9cm
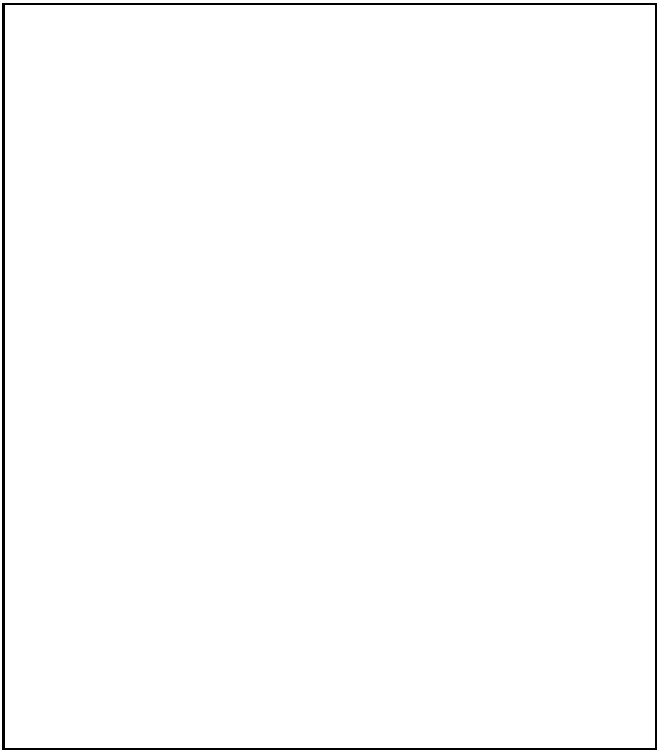
\figure{4.3}{Synthetic (V,V--I) CMD 
with a discretely increasing extinction up to A$_{\rm V}$\muspc=\muspc1\mag75
at 4 kpc distance
}
\endfig
%\newcolumn
%
\begfig 9.9cm
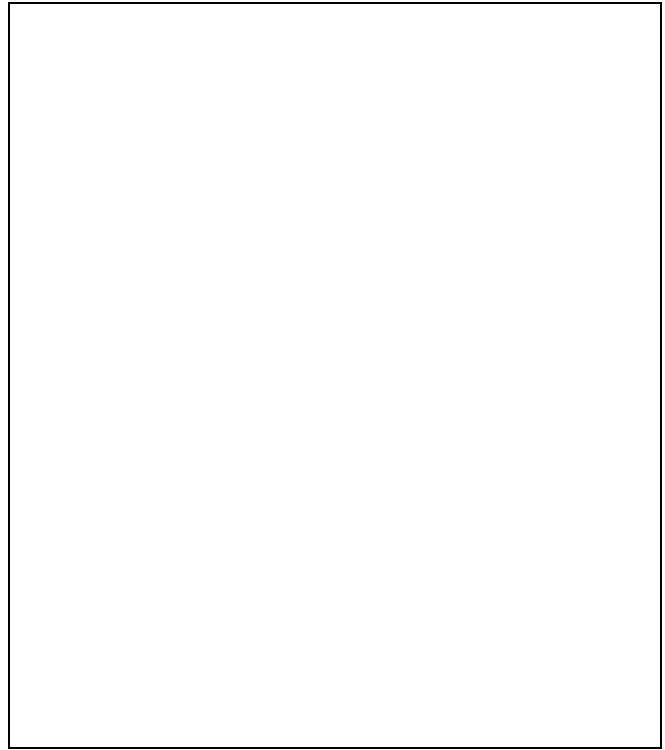
\figure{4.2}{Synthetic (V,V--I) CMD 
with a discretely increasing extinction up to A$_{\rm V}$\muspc=\muspc1\mag75
at 2 kpc distance
}
\endfig
The resolution
of the input, patchy extinction map is scaled dynamically 
in a \hbox{200\muspc$\times$\muspc200} grid between the minimum 
and maximum values.
A smooth map is obtained through an appropriate interpolation scheme 
between the neighbouring points.
The current choice of the figure dimensions results in an apparent,
in distance, extended extinction patch.
In the simulations 300 Poissonian distributed points are used 
as centers for these patches. Around these points another group of 
points are distributed within a radius~$\lambda$. This radius
actually defines how independent 
the patches are from each other.  
The patches give rise to
the threaded like substructures in Fig.~2c,
when they start to 
overlap with each other.
\par

\titleb{Results}
Figures 3.1\to3.3, 4.1\to4.3, 5.1\muspc\&\muspc5.2 and 6.1\muspc\&\muspc6.2 
show the resulting CMDs for the different 
type of extinction adopted. 
\par
Figures 3.1\to3.3 show the CMD when a linear 
extinction curve is used. These figures show the cases
where the extinction increases linearly, up 
to the maximum value at respectively 1, 2 and 4 kpc away
from us.
They further show the gradual tilt of the disc main sequence,
up to the distance where the extinction is maximum.
The sequence is vertical beyond that distance. 
\par
\noindent
Notice that the tilt of the disc main sequence has not the same 
orientation as the reddening vector.
This is due to differences in both the extinction and the 
distance of each star.
With an increasing distance for the maximum extinction,
the disc horizontal branch stars along the line of sight 
start to disperse, because they do not have the same extinction.
They also get a tilt, similar to the main sequence stars.
Figs. 1.1 and 3.3 are quite similar, this
is due to the fact that in first order the linear increasing 
extinction gives a good approximation to the actual extinction
(see Fig.~2a).
\par
\begfig 9.9cm
\figure{5.1}
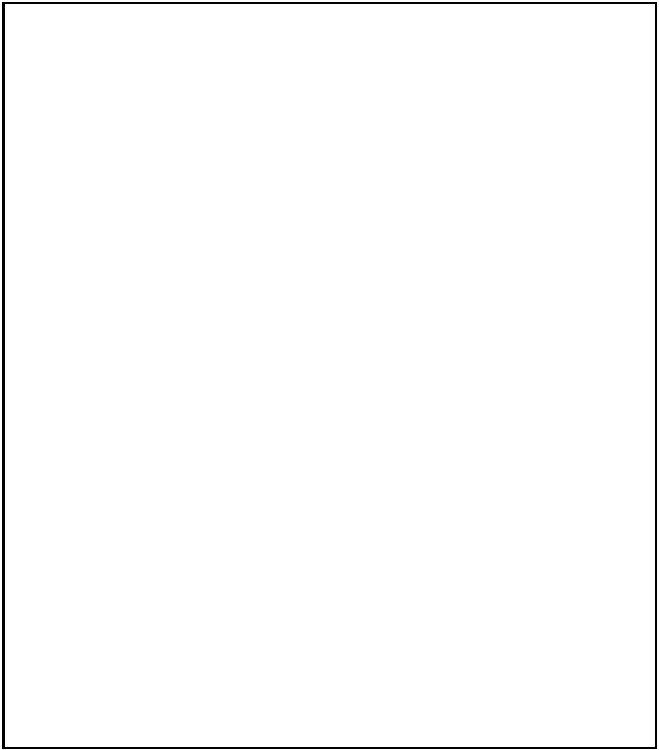
{Synthetic (V,V--I) CMD with a Poissonian distributed
extinction (A$_{\rm V}$\muspc=\muspc0\mag0\to1\mag75)}
\endfig
With a 4 kpc distance for maximum extinction and 
$b$\muspc=\muspc$-$\/4\fdg2 (Baade's Window, subfield \#3)
the outer boundary of this absorption layer is about 300 pc.
This is in agreement with an effective thickness of 100 pc
for an exponentially decreasing density distribution. 
It is emphasized that there is a discontinuity 
between the disc and the bar horizontal branches,
because we did not include the simulations of the older 
disc populations. But this is merely an artifact caused by
the simplifications made for these simulations.
The morphological structures of the bar population remain the
same, because all the stars are located outside the extinction
layer and all have a maximum extinction.
\par
Figures 4.1\to4.3 show the CMD when a step-like
extinction curve is used. They show the cases
for which the extinction increases suddenly, up 
to the maximum value at respectively 1, 2 and 4 kpc away
from us.
Due to the discrete step size of the extinction, the disc sequence
splits up in two separate vertical sequences: one bluer 
(V--I\muspc$\simeq$\muspc0\mag6)
and brighter than
the other (V--I\muspc$\simeq$\muspc1\mag25), 
because the brighter stars are located in front of the extinction
screen. 
Note that the two sequences are shifted from each other
along the reddening line.
The bright, blue sequence still connects with the 
fainter redder one through the lower mass main sequence stars
(the dots in the figures).
In contrast to the previous cases with a
linearly increasing extinction, there is now 
a clear separation between the evolved stars
from the disc and bar population.
In Fig. 4.3 one can clearly see the disc turn-off from the
stars which are located in front of the extinction screen. 
Also in these cases the
morphological structures from the bar population remain the
same, because all the stars are located behind the extinction
layer and all have a maximum extinction.
\par
\begfig 9.9cm
\figure{5.2}
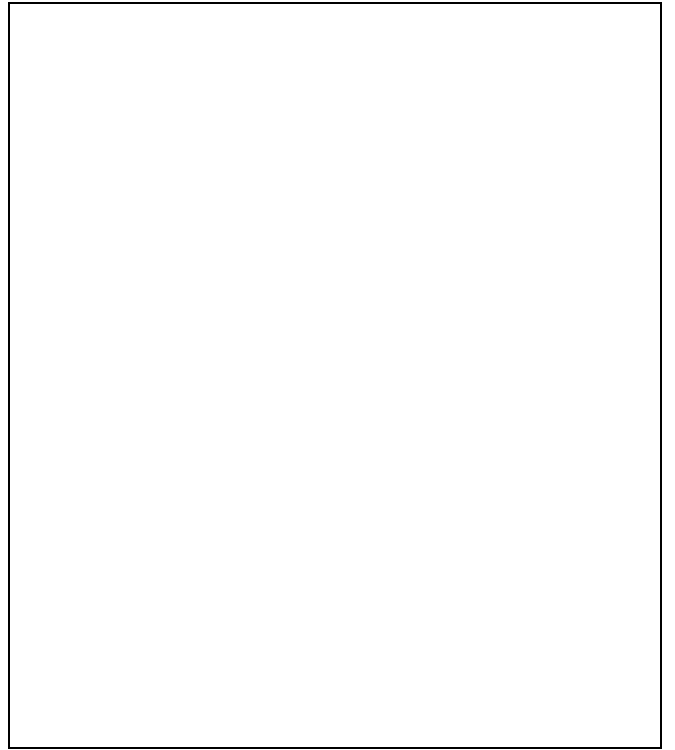
{Synthetic (V,V--I) CMD 
with a Poissonian distributed extinction 
(A$_{\rm V}$\muspc=\muspc1\mag60\muspc$\pm$\muspc0\mag20)
\vskip-0.48mm
}
\endfig
In general, one does not have merely a step-like increase of the extinction.
A combination with a gradual, with distance increasing extinction
is more realistic. Due to the very distinct nature of the 
step-like increase in the extinction, one can determine quite accurately
with an uncertainty of about 0.2~kpc
the distance to the source of this feature. Bertelli et al. (1995)
demonstrated this for the extinction of the field near the 
galactic cluster NGC~6603. The morphology of the disc sequence
in the CMD of the cluster Terzan~1 (Ortolani et al. 1993)
suggests that this is possibly another field, in which we can expect
a step-like increase for the extinction at a certain distance.
However, the analysis of the frame near this cluster 
(Bertelli et al. 1995)
did not reveal the presence of a steep increase in the extinction
along the line of sight. 
This probably is an other indication, that the extinction towards the 
galactic centre changes rapidly over a relatively small area. 
\par
Figures 5.1\muspc\&\muspc5.2 show the CMDs for Poissonian extinction.
In the first case A$_{\rm V}$ is varied between 0\mag0 and 1\mag75.
It is emphasized once more that this exercise is not representative
for the actual situation in Baade's Window.
The adopted boundary of the absorbing layer is located at about
4~kpc away from us and, with respect to Fig. 3.3, the extinction
can be lower at any given distance. 
The blue side of the disc structure will not be as tilted anymore
and is nearly vertical now.
Near the Galactic Centre some of the
stars will be brighter and bluer, which can be clearly seen in 
the horizontal branch of the bar population.
In Fig. 5.2 constraints are put to the lower
limit of the extinction variations. This case is quite similar
to Fig. 3.3, the only difference is that a dispersion is present
around the mean extinction.
This give rise to a little bit more dispersed morphology 
of the sequences in 
the synthetic CMD, but it will not introduce a change 
of the main morphological properties.
The mean extinction for the bar stars in this simulation
is lower than for Fig. 3.3.
This results in a small blueward shift (0\mag06) of 
the bar population in Fig. 5.2
with respect to Fig. 3.3.
\par
\begfig 9.9cm
\figure{6.1}
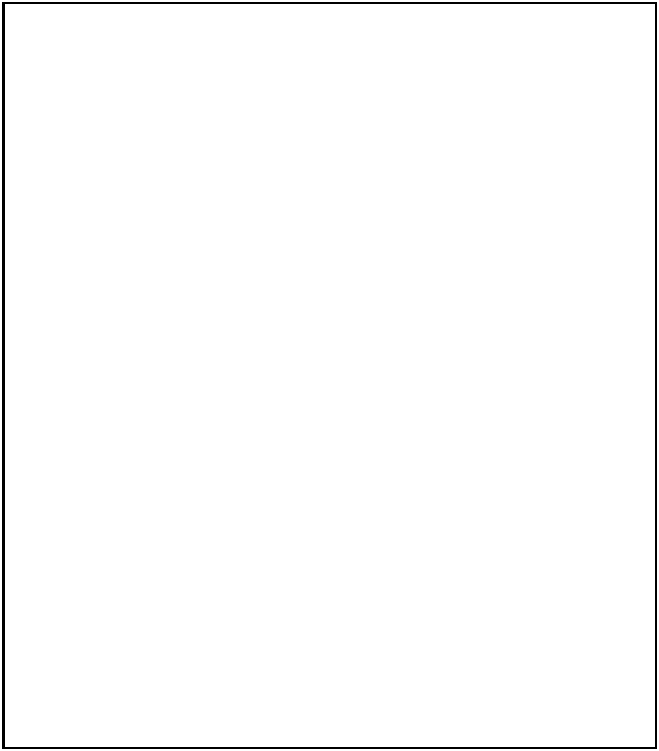
{Synthetic (V,V--I) CMD with a patchy
extinction ($\lambda$\muspc=\muspc0.3; 
\hbox{A$_{\rm V}$\muspc=\muspc1\mag60\muspc$\pm$\muspc0\mag20})}
\endfig
In the simulations of patchy extinction, 
A$_{\rm V}$ has been varied between 0\mag0 and 1\mag75
together with the correlation radius, respectively
\hbox{$\lambda$\muspc=\muspc0.3} and 0.6, between the patches.
The figures with \hbox{$\lambda$\muspc=\muspc0.3} and 0.6
are very similar. They differ in small
details concerning substructures in the morphology of the
different sequences in the CMD. 
Besides very small substructures,
the CMDs are very similar to the Poissonian case in Fig.~5.1
and they are therefore not displayed.
This result is not surprising, because all that matters is
the extinction range. The extinction patches only give
rise to some small substructures in the synthetic CMD. 
Similar arguments can be given for simulations with 
\hbox{A$_{\rm V}$\muspc=\muspc1\mag60\muspc$\pm$\muspc0\mag20}
and \hbox{$\lambda$\muspc=\muspc0.3} or 0.6 with respect to
Fig.~5.2. The differences between patchy and random extinction
are even smaller than in the previous case. 
Figure 6.1 shows the case \hbox{$\lambda$\muspc=\muspc0.3}.
For Fig.~6.2 an extinction range 
A$_{\rm V}$ between 1\mag3 and 2\mag3 is adopted. 
Again no big differences between the figures with
\hbox{$\lambda$\muspc=\muspc0.3} or 0.6
are noticeable.
We therefore consider only the case \hbox{$\lambda$\muspc=\muspc0.3}. 
Because of the extinction range,
this figure ought to be compared with the CMD from Baade's Window
for the whole field, see for example Fig~1.1 (Ng et al. 1996a).
Taking into account a 0\mag07 shift in (V--I) colour (Ng et al. 1996a),
partly due to differences in the zeropoint of the 
HRD-GST photometric system,
between the simulated and the observed CMD, the
following is noticed: the simulated RGB and HB both cover
the corresponding region in the observed CMD completely. 
\par
\begfig 9.9cm
\figure{6.2}
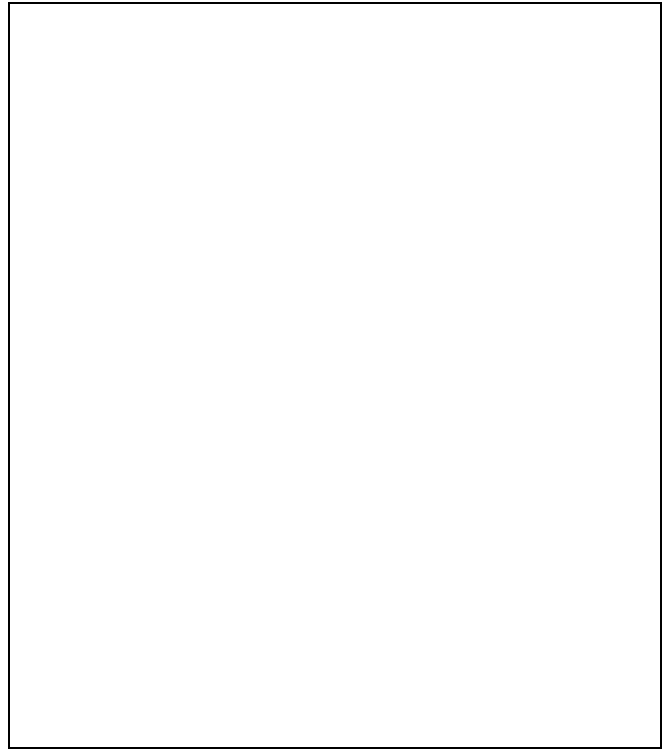
{Synthetic (V,V--I) CMD with a patchy extinction 
($\lambda$\muspc=\muspc0.3; 
\hbox{A$_{\rm V}$\muspc=\muspc1\mag80\muspc$\pm$\muspc0\mag50})}
\endfig
\noindent
This implies that \hbox{A$_{\rm V}$\muspc=\muspc1\mag3\to2\mag3}
covers indeed the extinction in Baade's Window. It also confirms 
that the stars from the `bar' population are dominating the  
RGB and HB morphological structures in the CMD. 
This does not imply that even metal-richer stars 
are absent in this field. It mainly shows that they
are not dominantly present in the CMD.
This is in agreement with the metallicity distribution
found from Baade's Window K-giants (Sadler et al. 1996).
Furthermore, the sequence due to the disc HB stars
becomes broader and actually constrains the maximum extinction
allowed along the line of sight. 
\par
The simulations in Figs.~6.1\muspc\&\muspc6.2 show that the actual choice 
of the correlation radius, besides the very small values, 
is likely not very important. On the other hand, if only a small 
number of patches had been used, the results might have shown 
some dependency with the choice of the value for the
correlation radius $\lambda$. But probably this is not 
a realistic option. Figures~6.1 \& 6.2 further demonstrates
that the HB morphology is due to  
differential reddening, combined with a metallicity
spread among the HB stars.
We are able to generate in the simulations a patchy like pattern for
the extinction along the line of sight. 
If the extinction difference between
the patchy `globules' is large enough and when these globules are
big enough, one of the questions that 
remains open for future analysis is: can we actually 
determine the distance to these patchy globules\/? 
In Baade's Window there is no big difference between a patchy
or a Poissonian type of extinction.
Only, when a large extinction range is present the difference
might become noticeable.
\par
\begfig 9.9cm
\figure{7}
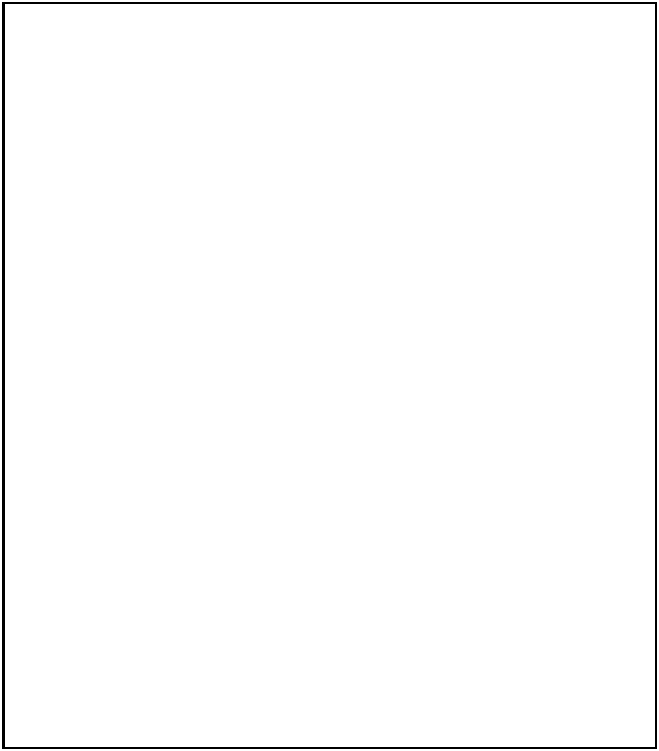
{Synthetic (J,J--K) CMD for Baade's Window. 
The absorption in the J and K passbands are
obtained under the assumption that $A_J/A_V=0.282$
and $A_K/A_V=0.112$ (Rieke \& Lebofsky 1985),
with $A_V$\muspc=\muspc1\mag70.
}
\endfig

\titlea{Discussion}
MC simulations have been made for various types of extinction.
The main extinction is caused
by material inside an absorbing layer. 
In first order the boundary of this layer is at 
\hbox{$z$\muspc$\simeq$\muspc300~pc}.
The stellar contribution in this layer is due to disc stars from 
various populations along the line of sight. 
Any significant change in the extinction will show up clearly in the
CMD morphology of the foreground disc sequence. 
But towards the galactic centre and Baade's Window
this will only hold up to approximately a 5~kpc distance from
the Sun, where a significant decrease in the disc density appears 
to be present (Bertelli et al. 1995; Ng et al. 1996a).
For practical purposes one might assume a constant extinction
beyond 5~kpc. On the other hand, the bari-centre of the 
bulge/bar red HB sequence provides information of the extinction
at 8~kpc distance. 
A suitable interpolation, between this and the extinction at about 
5~kpc, results in the extinction curves obtained
in star counts studies with the HRD-GST. 
\par
If the red horizontal branch stars in Baade's Window is due to 
stars with a mean solar metallicity one has
to disperse the horizontal branch stars along 
the reddening line and introduce
a dispersion in the extinction, i.e. differential reddening.
In that case, a smaller dispersion will be also present
in the foreground extinction.
If the dispersion is too large the disc sequence will
change significantly. On the other hand, a small 
extinction dispersion, which is comparable with the  
colour dispersion of the disc sequence, will not give rise
to a significant change in both the disc and bar morphology.
\par
Patchiness of the extinction might be such,
that it may introduce a change in the CMD the horizontal branch 
morphology of a population which has a mean solar metallicity, without
influencing significantly the disc sequence.
But this will require very special conditions for the patchiness
of the extinction.
One can imagine that these requirements can be found in a certain direction,
but they will not be ubiquitous. 
An inspection of the CMDs from the 
various subfields in and near Baade's Window
(Udalski et al. 1993), some offset fields near globular
clusters (Ortolani et al. 1990, Bica et al. 1994) and
the CMDs of metal-rich globular clusters
(Ortolani et al. 1990, 1992, 1993, 1994, 1995b \& 1996)
shows that differences in the (mean) extinction are
certainly present. \par
The horizontal branches from the
field and the clusters all seem share something in common.
It is unlikely that the change in the extinction
is always such that the disc sequence in the CMDs is not influenced
significantly, while the horizontal branches are dispersed
along the direction of the reddening line.
In fact the morphology of the disc sequence is always affected 
first by any change in the reddening.
It is more likely that we are dealing with an underlying population
which has, due to primarily a metallicity spread, 
a horizontal branch almost parallel to the extinction line.
This population, such as the `bar' population identified by Ng et al.
(1996a\&c), will not require special patchy extinction conditions.
Ng et al. (1996d) demonstrated for metal-rich globular clusters
that a stellar population similar to
the `bar' population can give a proper description of the cluster HBs. 
But one ought to be cautious with clusters at distances near
8~kpc. In those cases the cluster HB-clump overlaps 
with the bulge/bar clump and it will be difficult
to separate the two contributions from each other (Ng et al. 1996d).
\par
If one insists on a small metallicity range, a larger extinction range,
\hbox{$\Delta$A$_V$\muspc$\simeq$\muspc0\mag7},
will be required along the line of sight, in order to obtain a
comparable morphological appearance for the cluster HBs.
Such an increase in the extinction range will certainly affect the
morphology of the disc sequence. 
The current analysis allows a 
change in the extinction of 
\hbox{$\Delta$A$_V$\muspc$\simeq$\muspc0\mag2\to0\mag3}.
Such a value invokes only marginal changes in the morphology
of the disc sequence in Figs.~6.1
and 6.2.
Merely differential reddening or a large metallicity range 
will not be sufficient to explain the red HB morphology.
This constraint therefore favours a large metallicity range
combined with differential reddening. 
\par
In (V,V--I) CMDs age-metallicity and
extinction effects cannot be separated unambiguously
from each other when a large metallicity spread is present. 
Observations in other near infrared 
passbands are needed. In a (J,J--K) CMD extinction effects 
and metallicity effects are not pointing in the same direction.
A simulated (J,J--K) CMD in Fig.~7 indicates that
a horizontal spread in colour of the red horizontal branch stars
is likely due to metallicity effects, while the effects 
of extinction will give a 
dispersion in another direction.
Furthermore, the morphology of the (K,J--K) CMD of NGC 6528 
from Guarnieri et al. (1995) indicates that this cluster 
probably has an almost constant extinction. The 
colour dispersion of the horizontal branch stars
is likely due to age-metallicity effects.
On the other hand, the large near infrared sky surveys
(DENIS, Epchtein et al. 1993; 2MASS, Kleinmann 1992)
might be suitable to study the variations of the 
extinction in large areas. Thus providing stronger
constraints for the determination of the ages and metallicities
of the stellar populations towards the 
Galactic Centre.
The simulated (J,J--K) CMD, displayed in Fig.~7, shows 
that a study like this is feasible, even with 
\hbox{K$_{lim}$\muspc=\muspc14\mag5}. 
\par 
Simulations have been presented for different types of extinction.
These simulations show that distinct differences are present 
between the morphology of the disc sequences for the different
forms of extinction. 
They further show that the disc sequence is a good indicator
for the nearby extinction along the line of sight.
The difference between patchy and Poissonian
extinction is not significant enough, in order to explain 
the HB-clump morphology 
in Baade's Window solely by patchy, differential extinction.
Differential extinction combined with a metallicity spread among 
the HB stars is favoured as explanation for the red HB morphology.
\par
\bigskip
\acknow{Y.K. Ng is indebted to R. van de Weygaert for providing 
an algorithm to generate correlated distributions. 
Once more I made good use of it by generating patchy extinction
maps.
K.Z. Stanek is acknowledged for his constructive comments.
Y.K. Ng thanks the Padova Astronomical Department 
for its hospitality. 
ANTARES, an astrophysics network funded by the HCM programme of the
European community, provided financial support for Ng's research visit
to Padova. Ng is supported by HCM grant CHRX-CT94-0627
from the European Community.
G. Bertelli acknowledges the financial
support received from the Italian Ministry of University, Scientific 
Research and Technology (MURST). 
}
\par

\hyphenation{Nijmegen}
\begref{References}
\ref Arp H., 1965, ApJ 141, 43
\ref Bertelli G., Bressan A., Chiosi C., Fagotto F., Nasi E., 
1994a, A\&AS 106, 275
\ref Bertelli G., Bressan A., Chiosi C., Ng Y.K., Ortolani S.,
     1994b, Mem.S.A.It. 65, 689
\ref Bertelli G., Bressan A., Chiosi C., Ng Y.K., Ortolani S.,
     1995, A\&A 301, 381 (Paper II)
\ref Bertelli G., Bressan A., Chiosi C., Ng Y.K., 
     1996, A\&A 310, 115
%\ref Bertelli G., Ng Y.K., Chiosi C., Bressan A.,
%     1996b, {\it in preparation} 
\ref Bica E., Ortolani S., Barbuy B., 1994, A\&A 283, 67
\ref Catalan M., de Freitas Pacheco J.A., 1996, PASP 108, 166
\ref Cox D.P., Proceedings IAU Symposium 120,
     {\it `Structure and Dynamics of the Interstellar Medium'},
     17\to21 April 1989, Granada (Spain),
     G Tenorio-Tagle, M. Moles and J. Melnick (eds.), 500
\ref Epchtein N., et al., 1993, DENIS bluebook
\ref Guarnieri M.D., Montegriffo P., Ortolani S., 
Moneti A., Barbuy B., Bica E., 1995, The Messenger 79, 26
\ref Jones T.J., Ashley M., Hyland A.R., Ruelas-Mayorga A.,
1981, MNRAS 197, 413
\ref Kleinmann S.G., 1992, Proceedings 
     {\it ` Robotic telescopes in the 1990s'}
     103rd Annual Meeting of the
     Astronomical Society of the Pacific, Univ. of
     Wyoming, Laramie, June 22-24, 1991, 
A.V. Filippenko (ed.), ASP Conference Series 34, 203
\ref Ng Y.K., 1994, Ph.D. thesis, Leiden University, the Netherlands
\ref Ng Y.K., 1996, Proceedings
     `The impact of large-scale near-IR sky surveys',
     24\to26 April 1996, Puerto de la Cruz (Tenerife; Spain),
     P. Garzon-lopez (ed.), {\it in press}
\ref Ng Y.K., Bertelli G., Bressan A., Chiosi C., Lub J., 1995,
     A\&A 295, 655 (Paper I; erratum A\&A 301, 318)
\ref Ng Y.K., Bertelli G., Chiosi C., Bressan A., 1996a
     A\&A 310, 771 (Paper III)
\ref Ng Y.K., Bertelli G., Chiosi C., Bressan A., 1996b,
     A\&A {\it submitted} 
\ref Ng Y.K., Bertelli G., Chiosi C., Bressan A., 1996c,
     Workshop `Spiral Galaxies in the near IR',
     Garching bei M\"unchen, 7--9 June 1995, D. Minniti and
     H.-W. Rix (eds.), 110
\ref Ng Y.K., Bertelli G., Chiosi C., 1996d, A\&A {\it submitted} 
%\ref Ng Y.K., Bertelli G., Chiosi C., 1996e, A\&A {\it to be submitted} 
\ref Ortolani S., Barbuy B., Bica E., 1990, A\&A 236, 362
\ref Ortolani S., Bica E., Barbuy B., 1992, A\&AS 92, 441 
\ref Ortolani S., Bica E., Barbuy B., 1993, A\&A 267, 66 
\ref Ortolani S., Barbuy B., Bica E., 1994, A\&AS 108, 653 
\ref Ortolani S., Renzini A., Gilmozzi R., et al., 
%Marconi, G., Barbuy, B., Bica, E., Rich, R.M., 
     1995a, Nature 377, 701
\ref Ortolani S., Barbuy B., Bica E., 1995b, Messenger 82, 10
\ref Ortolani S., Barbuy B., Bica E., 1996, A\&A 308, 733 
\ref Paczy\'nski B., Stanek K.Z., Udalski A., et al.,
%Szyma\'nski, M., Ka{\l}u{\zdot}ny, J., Kubiak, M., Mateo, M., 
     1994, AJ 107, 2060
\ref Renzini A., 1996, workshop `Spiral Galaxies in the near IR',
     Garching bei M\"unchen, 7--9 June 1995, D. Minniti and 
     H.-W. Rix (eds.), 95
\ref Rieke G.H., Lebofsky M.J., 1985, ApJ 288, 618
\ref Sadler E.M., Rich R.M., Terndrup D.M., 1996, AJ {\it in press}
\ref Stanek K.Z., 1996, ApJ 460, L37
\ref Ruelas-Mayorga R.A., Rev. Mex. Astron. Astrofis. 22, 27
\ref Udalski A., Szyma\'nski M., Ka{\l}u{\zdot}ny J., Kubiak M.,
     Mateo M., 1993, Acta Astron. 43, 69
\ref Wesselink Th.J.H., 1987, Ph.D. thesis, Catholic University
of Nijmegen, the Netherlands
\ref Wo\'zniak P.R., Stanek K.Z., 1996, ApJ 464, 233
\endref

\bye